\documentclass[prd,nofootinbib,showpacs,preprintnumbers,amsmath,amssymb]{revtex4}

\usepackage{graphicx}
\usepackage{dcolumn}
\usepackage{bm}

\def\bear{\begin{eqnarray}}
\def\ear{\end{eqnarray}}

\begin{document}

\begin{flushright}
YITP-13-61
\end{flushright}

\title{Geometric Origin of Stokes Phenomenon for de Sitter Radiation}

\author{Sang Pyo Kim}\email{sangkim@kunsan.ac.kr}
\affiliation{Department of Physics, Kunsan National University, Kunsan 573-701, Korea\footnote{Permanent address}}
\affiliation{Yukawa Institute for Theoretical Physics, Kyoto University, Kyoto 606-8502, Japan}

\begin{abstract}
We propose a geometric interpretation for the Stokes phenomenon in de Sitter spacetime that particles are produced in even dimensions but not in odd dimensions. The scattering amplitude square for a quantum field between the in-vacuum and the transported one along a closed path in
the complex-time plane gives the particle-production rate that explains not only the Boltzmann factor from the simple pole at infinity, corresponding to the cosmological horizon, but also the sinusoidal behavior from simple poles at the north and south poles of the Euclidean geometry. The Stokes phenomenon is a consequence of interference among four independent closed paths in the complex plane.
\end{abstract}

\date{\today}
\pacs{04.60.-m, 04.62.+v, 03.65.Vf, 03.65.Sq}

\maketitle

\section{Introduction} \label{sec I}

Nature in some circumstances distinguishes the dimensionality of spacetime through the underlying theory. The fundamental solution to the wave equation has a delta-function support in even dimensions while it has a step-function support in odd dimensions. Another interesting feature of dimensionality is that a de Sitter (dS) spacetime produces particles in even dimensions while it does not produce any particles in odd dimensions \cite{Mottola,BMS}. Polyakov interpreted the reflectionless scattering of a quantum field in an odd-dimensional dS spacetime as a soliton of the Korteweg-deVries (KdV) equation \cite{Polyakov08,Polyakov10}. In fact, the quantum field in all odd-dimensional dS spacetimes has the P\"{o}sch-Teller potential \cite{DrazinJohnson}, whose asymptotically reflectionless scattering implies the solitonic nature of produced particles \cite{Kim11}.

In the phase-integral method \cite{FromanFroman} each mode of a quantum field in time-dependent gauge fields or curved spacetimes has at least one pair of complex turning points in the complex-time plane, for which the Hamilton-Jacobi action between each pair determines the particle-production rate for that channel. Remarkably the actions from more than one pair of complex turning points may contribute constructively and destructively to the particle production, known as the Stokes phenomenon. The Stokes phenomenon in Schwinger mechanism discovered by Dumlu and Dunne explains the substructure of produced particles for some time-dependent electric fields \cite{DumluDunne10,DumluDunne11,DumluDunne11-2}. In a dS spacetime a quantum field has one pair of complex turning points in the planar coordinates and two pairs in the global coordinates. The actions along Stokes lines connecting two anti-Stokes lines have both the imaginary part determining the dS radiation and the real part resulting in constructive or destructive interference for dS radiation in the global coordinates \cite{Kim10}.

In this paper we propose a geometric interpretation in the complex-time plane of the Stokes phenomenon for dS radiation. Instead of tunneling paths and their actions in the phase-integral method, we study the quantum evolution operator for the field and calculate the geometric contributions to the transported in-vacuum in the complex-time plane. It is shown that each harmonics of the field in the global coordinates of dS spacetime obtains not only a geometric factor for dS radiation which originates from the simple pole at infinity corresponding to the cosmological horizon but also interfering terms from finite simple poles at the north and south poles of the Euclidean geometry which explain the sinusoidal behavior responsible for the presence or absence of particle production in even and odd dimensions.

The geometric transition of the time-dependent Hamiltonian in the complex plane leads to an exponential decay of the initial state through level crossings
\cite{HwangPechukas,JKP}. The geometric transition has been formulated to include the higher corrections in Ref. \cite{KKS}. Recently it has been shown that the in-vacuum of a time-dependent oscillator transported along a closed path in the complex plane may gain a geometric contribution from the simple pole at infinity and that the geometric factor explains Schwinger pair production in a constant electric field and dS radiation in the planar coordinates \cite{Kim13}.
It has been further argued that the scattering amplitude between the transported in-vacuum and the in-vacuum determines multiple pair production, depending on the winding number of the closed path in the complex plane. However, these models have only one pair of complex turning points and rule out the Stokes phenomenon, whereas the global coordinates of dS spacetime provide two pairs of complex turning points for each harmonics of quantum field and lead to the Stokes phenomenon.

The organization of this paper is as follows. In Sec. \ref{sec II} the real-time evolution of a quantum field is formulated in the functional Schr\"{o}dinger picture. In Sec. \ref{sec III} the scattering amplitudes between the in-vacuum and the transported one along closed paths in the complex-time plane are computed
and the particle-production rate is given by summing the scattering amplitude squares for all independent paths of winding number one. It is shown that the Stokes phenomenon is a consequence of the interference among different paths, which has a geometric origin. In Sec. \ref{sec IV} we compare the result of this paper with other methods and discuss the physical implications.

\section{Evolution Operator in Real Time} \label{sec II}

For the sake of simple harmonics decomposition we consider a complex scalar in the global coordinates of a (d+1)-dimensional  dS spacetime (in units of $c = \hbar =1$)
\begin{eqnarray}
ds^2 = -dt^2 + \frac{1}{H^2} \cosh^2 (Ht) d \Omega_d^2. \label{metric}
\end{eqnarray}
The field equation for the complex scalar field with mass $m$ may be derived from the Lagrangian\footnote{The Lagrangian under a field redefinition $\bar{\psi} = \psi/(-g)^{1/4}$ is equivalent to that of Ref. \cite{BirrellDavies}.}
\begin{eqnarray}
L_{\phi} (t) = \int \sqrt{-g} d^dx \frac{1}{2} \Bigl( \psi^* \square \psi - m^2 \psi^* \psi \Bigr),
\end{eqnarray}
where $\square =  (1/\sqrt{-g}) \partial_{\mu} (\sqrt{-g} g^{\mu \nu} \partial_{\nu})$. Decomposing $\psi$ and $\psi^*$ by the spherical harmonics on $S^d$,  $\nabla^2 u_{\kappa} ({\bf x}) = - \kappa^2 u_{\kappa} ({\bf x})$  with $\kappa^2 = l (l+ d-1), (l = 0, 1, \cdots )$ \cite{RubinOrdonez} and symmetrizing them, we obtain the Hamiltonian
\begin{eqnarray}
H_{\phi} (t) = \sum_{\kappa} \frac{1}{2} \Bigl[\Bigl(\pi_{\kappa}^* + \frac{\dot{g}}{4g} \psi_{\kappa} \Bigr) \Bigl(\pi_{\kappa} + \frac{\dot{g}}{4g} \psi_{\kappa}^* \Bigr)+ \omega_{\kappa}^2 (t) \psi^*_{\kappa} \psi_{\kappa}  \Bigr], \label{ham}
\end{eqnarray}
where $\pi_{\kappa} = \dot{\psi}^*_{\kappa} + (\dot{g}/4g)\psi^*_{\kappa}$ and $\pi^*_{\kappa} = \dot{\psi}_{\kappa} + (\dot{g}/4g)\psi_{\kappa}$, and
\begin{eqnarray}
\omega_{\kappa}^2 (t) = \gamma^2 + \frac{(\lambda H)^2}{\cosh^2 (Ht)}, \label{freq}
\end{eqnarray}
where for a massive scalar $(m > dH/2)$
\begin{eqnarray}
\gamma = \sqrt{m^2 - \frac{(dH)^2}{4}}, \quad \lambda = \sqrt{\kappa^2 + \frac{d^2}{4}}.
\end{eqnarray}

In the functional Schr\"{o}dinger picture, the evolution operator for the field obeys the time-dependent Schr\"{o}dinger equation
\begin{eqnarray}
i \frac{\partial}{\partial t} \prod_{\kappa} \hat{U}_{\kappa} (t) = \sum_{\kappa} \hat{H}_{\kappa} (t) \prod_{\kappa} \hat{U}_{\kappa} (t). \label{sch eq}
\end{eqnarray}
Each Hamiltonian is diagonalized by the time-dependent annihilation and creation operators as
\begin{eqnarray}
\hat{H}_{\kappa} (t) = \omega_{\kappa} (t) \Bigl(\hat{a}^{\dagger}_{\kappa} (t) \hat{a}_{\kappa} (t) + \frac{1}{2} \Bigr). \label{num ham}
\end{eqnarray}
Then the evolution operator is expressed by the spectral resolution \cite{Kim13}
\begin{eqnarray}
\hat{U}_{\kappa} (t, t_0) = \Phi_{\kappa}^T (t) {\rm T} \exp \bigl[- i \int_{t_0}^{t} \bigl(
{\bf H}_{\kappa D} (t') - {\bf A}_{\kappa}^T (t') \bigr) dt' \bigr] \Phi_{\kappa}^*(t_0), \label{ev op}
\end{eqnarray}
where ${\bf H}_{\kappa D} (t) = \omega_{\kappa} (t)\,{\rm diag}\,(1/2, \cdots, n+1/2, \cdots)$ is the diagonal matrix
and $\Phi^T(t) = ( \vert 0_{\kappa}, t \rangle , \cdots,  \vert n_{\kappa}, t \rangle, \cdots)$ is the row vector of the number states for (\ref{num ham}),
and ${\bf A}_{\kappa}(t)$ is the induced vector potential from the time-dependent number states with entries
\begin{eqnarray}
\bigl( {\bf A}_{\kappa} (t) \bigr)_{mn}  = i \langle m, t \vert (\frac{\partial}{\partial t} \vert n, t \rangle) =
 i \frac{\dot{\omega}_{\kappa} (t)}{4 \omega_{\kappa} (t)} \bigl(\sqrt{n(n-1)}\delta_{m n-2} - \sqrt{(n+1)(n+2)} \delta_{m n+2} \bigr).
\end{eqnarray}
Here $T$ denotes the transpose of the matrix or vector.  Note that $\omega_{\kappa} (t) > 0$ and ${\bf A}_{\kappa} (t)$ does not have any singularity, so $\hat{U}_{\kappa} (t_0, t_0) = I$ for any path along the real-time axis and the in-in formalism thus becomes trivial.
In the real-time dynamics the in-out formalism carries all physical information through the scattering matrix between the out-vacuum and the in-vacuum.
Hence, to implement the in-in formalism for particle production, the real-time dynamics should be extended to the complex-time plane, as will be shown in the next section.

\section{Geometric Interpretation of Stokes Phenomenon} \label{sec III}

It has been known for long that the quantum evolution of a time-dependent Hamiltonian system exhibits a rich structure in the complex-time plane, such as geometric phases and nonadiabatic evolutions \cite{HwangPechukas,JKP,ShapereWilzcek}. Now we extend the quantum evolution (\ref{sch eq}) to a complex plane. For that purpose we assume that the geometry  (\ref{metric}) and the Hamiltonian (\ref{ham}) have an analytical continuation in
the whole complex plane, which is realized by the conformal mapping
\begin{eqnarray}
e^{Ht} = z, \quad (- \pi < {\rm arg}\, (Ht) \leq \pi).
\end{eqnarray}
Being interested in the quantum evolution along a path $z(t)$ in the complex plane, we analytically continue Eq. (\ref{sch eq}) to
\begin{eqnarray}
i \frac{\partial}{\partial z} \prod_{\kappa} \hat{U}_{\kappa} (z) = \sum_{\kappa}  \hat{H}_{\kappa} (z) \prod_{\kappa} \hat{U}_{\kappa} (z), \label{com ev}
\end{eqnarray}
where $\hat{H}_{\kappa} (z) := \frac{\partial t}{\partial z}  \hat{H}_{\kappa} (t(z))$. The quantum field theory in analytically
continued geometries has also been discussed in Ref. \cite{Kim99}.

Hence the spectrally resolved evolution operator (\ref{ev op}) is analytically continued to a closed path $C(z)$ in the complex plane provided that $\langle m, z \vert n, z \rangle$ holds along the path. Then the lowest order of the Magnus expansion \cite{Magnus,BCOR} gives the scattering amplitude between the in-vacuum and the transported one along a path $C^{(n)}$ of winding
number $n$ with the base point $t_0$ \cite{Kim13}
\begin{eqnarray}
\langle 0_{\kappa}, t_0 \vert 0_{\kappa}, C^{(n)} (t_0) \rangle = \exp \Bigl[- \frac{i}{2} \oint_{C^{(n)} (t_0)} \omega(z)dz \Bigr]. \label{sc am}
\end{eqnarray}
In the in-out formalism the vacuum persistence, which is the magnitude of the square of the scattering amplitude between the out-vacuum and the in-vacuum, is the probability for the out-vacuum to remain in the in-vacuum. The decay of the vacuum persistence results from one-pair and multipair production in bosonic theory \cite{KLY08}. Similarly, the magnitude of the scattering amplitude square (\ref{sc am}) is the rate for multiparticle production
\begin{eqnarray}
{\cal N}_{\kappa}^{(n)} = \Bigl\vert \exp \Bigl[- i \oint_{C^{(n)} (t_0)} \omega(z)dz \Bigr] \Bigr\vert,
\end{eqnarray}
and depends only on the information of simple poles included in the path.
The pair-production rate in time-dependent electric fields has been proposed of the form $\vert e^{- i \oint_{C^{(1)}}  \omega(z)dz} \vert$  in the
phase-integral method \cite{KimPage07}. The dynamical phase has an extension to the complex plane along $C^{(n)} (t_0)$ as
\begin{eqnarray}
\oint \omega_{\kappa} (t) dt = \frac{\gamma}{H} \oint \frac{1}{(z^2 + 1)z} \Bigl((z-z_+^*)(z-z_+)(z- z_-^*)(z-z_-) \Bigr)^{1/2} dz, \label{omega}
\end{eqnarray}
where the branch points are
\begin{eqnarray}
z_+ = \Bigl(\sqrt{1+ \frac{(\lambda H)^2}{\gamma^2}} + \frac{\lambda H}{\gamma}  \Bigr) e^{i \frac{\pi}{2}}, \quad z_- = \Bigl(\sqrt{1+ \frac{(\lambda H)^2}{\gamma^2}} - \frac{\lambda H}{\gamma} \Bigr) e^{i \frac{\pi}{2}}.
\end{eqnarray}
Cutting the branch points $z_{+}$, $z_{-}$ and their conjugates $z_{+}^*$, $z_{-}^*$ as shown in Fig. 1, the integrand in Eq. (\ref{omega}) is an analytic
function. The integrand (\ref{omega}) has a simple pole at $z = \infty$, which corresponds to the cosmological horizon, and
which is located outside the path and can be obtained by the large $z$ expansion \cite{Markushevich}. The geometric contribution from the pole at infinity is universal for all paths of nonzero winding numbers. Further, there are two finite simple poles at $z_i = i$ and $z_i = -i$, which correspond to the north and south poles of the Euclidean geometry of dS space (\ref{metric}).

The simple poles at $z = \pm i$ classify four independent paths of winding number 1 with the base point $t_0$  in the $z$ plane: the first class $C_{I}^{(1)}$ does not include any pole at $z = \pm i$ as shown in Fig. 1, the second class $C_{II}^{(1)}$ and the third class $C_{III}^{(1)}$ include only one pole at  $z = \pm i$ as shown in the left panel of Fig. 2, and the fourth class $C_{IV}^{(1)}$ includes both poles at $z = \pm i$ as shown in the right panel of Fig. 2. The scattering amplitude between the in-vacuum and the transported one along a path of each class always receives a geometric contribution $-2 i \pi {\rm Res}\, \omega (\infty)$ from the simple pole at $z = \infty$, which is located outside the path. The particle-production rate is the magnitude of the sum of the scattering amplitude square for each class path
\begin{eqnarray}
{\cal N}_{\kappa} = \Bigl\vert \sum_{J = 1}^{4} \langle 0_{\kappa}, t_0 \vert 0_{\kappa}, C_{J}^{(1)} (t_0) \rangle^2 \Bigl\vert
= \vert (1+ 2 e^{2 i \pi \lambda} + e^{4 i \pi \lambda})\vert e^{- 2 \pi \frac{\gamma}{H}}.
\end{eqnarray}
Here the first term in the parenthesis comes from $C_{I}^{(1)} (t_0)$, the second term from $C_{II}^{(1)} (t_0)$ and $C_{III}^{(1)} (t_0)$, and the last term from $C_{IV}^{(1)} (t_0)$. It should be noted that the magnitude is taken after summing over all independent paths of winding number 1.
In the limit of large action ($|\oint \omega_{\kappa}| \gg 1$ and $l \gg 1$), we approximately have $\lambda \approx l+d/2 - 1/2$ and obtain the particle-production rate
\begin{eqnarray}
{\cal N}_{\kappa}  = 4 \sin^2 \bigl(\pi (l+ d/2 )\bigr) e^{- 2 \pi \frac{\gamma}{H}}.
\end{eqnarray}
Hence, in odd dimensions ($d$ even) the particle-production rate vanishes while in even dimensions it is the leading Boltzmann factor of the exact formula \cite{BMS}
\begin{eqnarray}
{\cal N}_{\kappa} = \frac{\sin^2 \bigl(\pi (l+ d/2 )\bigr)}{\sinh^2 (\pi \gamma/H)}.
\end{eqnarray}

Finally, we compare the result of this paper with the Stokes phenomenon in the phase-integral method \cite{Kim10}.
In the complex plane $\{z_{-}^*, z_{-} \}$ and $\{z_{+}, z_{+}^* \}$ constitute two pairs of complex turning points and
each pair gives the Hamilton-Jacobi action for the scattering over barrier
\begin{eqnarray}
2 \int_{z_{-}^*}^{z_{-}} \omega_{\kappa} (z) dz = 2 \int_{z_{+}}^{z_{+}^*} \omega_{\kappa} (z) dz= \oint_{C^{(1)}_{J}} \omega_{\kappa} (z)dz = - 2 i \pi \frac{\gamma}{H} - 2 \pi \lambda,
\end{eqnarray}
where $J$ denotes the class $II$ or $III$. The imaginary and real parts of the actions determine the exponential and oscillatory behaviors for the particle-production rate, respectively \cite{Kim10}.
Thus the Stokes phenomenon for dS radiation originates from the interference among four independent paths involving two simple poles at the north and south poles of the Euclidean geometry.

\begin{figure}[t]
{\includegraphics[width=0.4\linewidth]{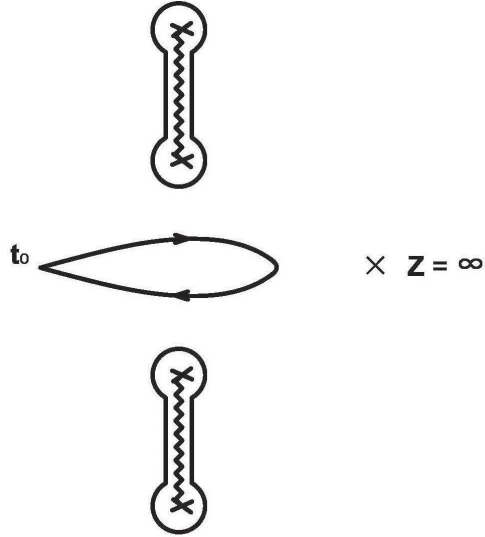}}
\caption{A pair of branch points $Z_+$ and $Z_-$ is cut by a line segment in the upper half of the plane and another pair $Z_+^*$ and $Z_-^*$ is cut by another line segment in the lower half of the plane. The first class consists of closed paths $C_{I}^{(1)}$ of winding number 1 that start from an initial time $t_0$ and do not include any finite simple poles. But the path still receives a geometric factor $-2 i \pi {\rm Res}\, \omega (\infty)$ from the simple pole at the infinity.} \label{contour-1}
\end{figure}

\begin{figure}[t]
{\includegraphics[width=0.4\linewidth]{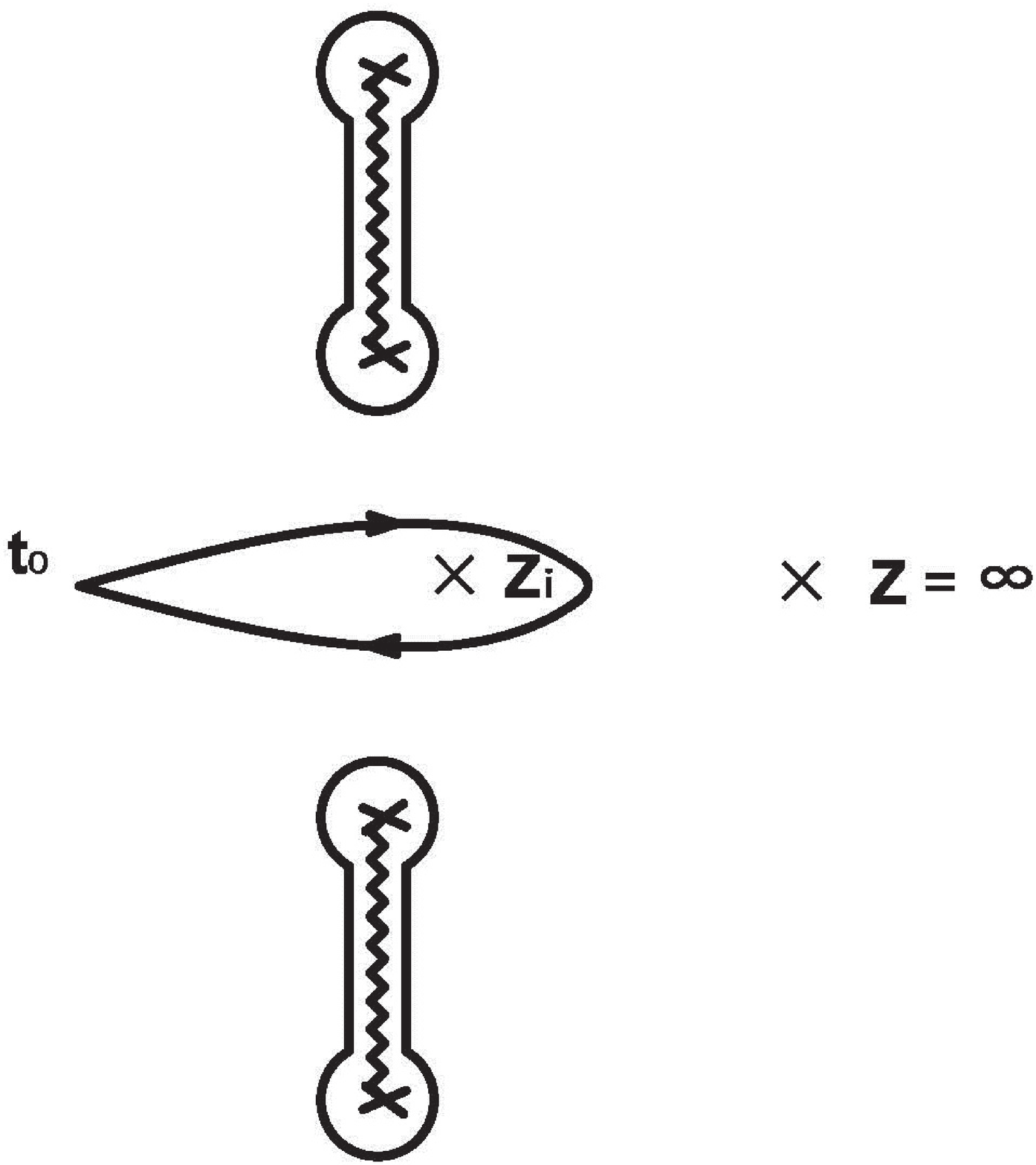}} \hfill
{\includegraphics[width=0.4\linewidth]{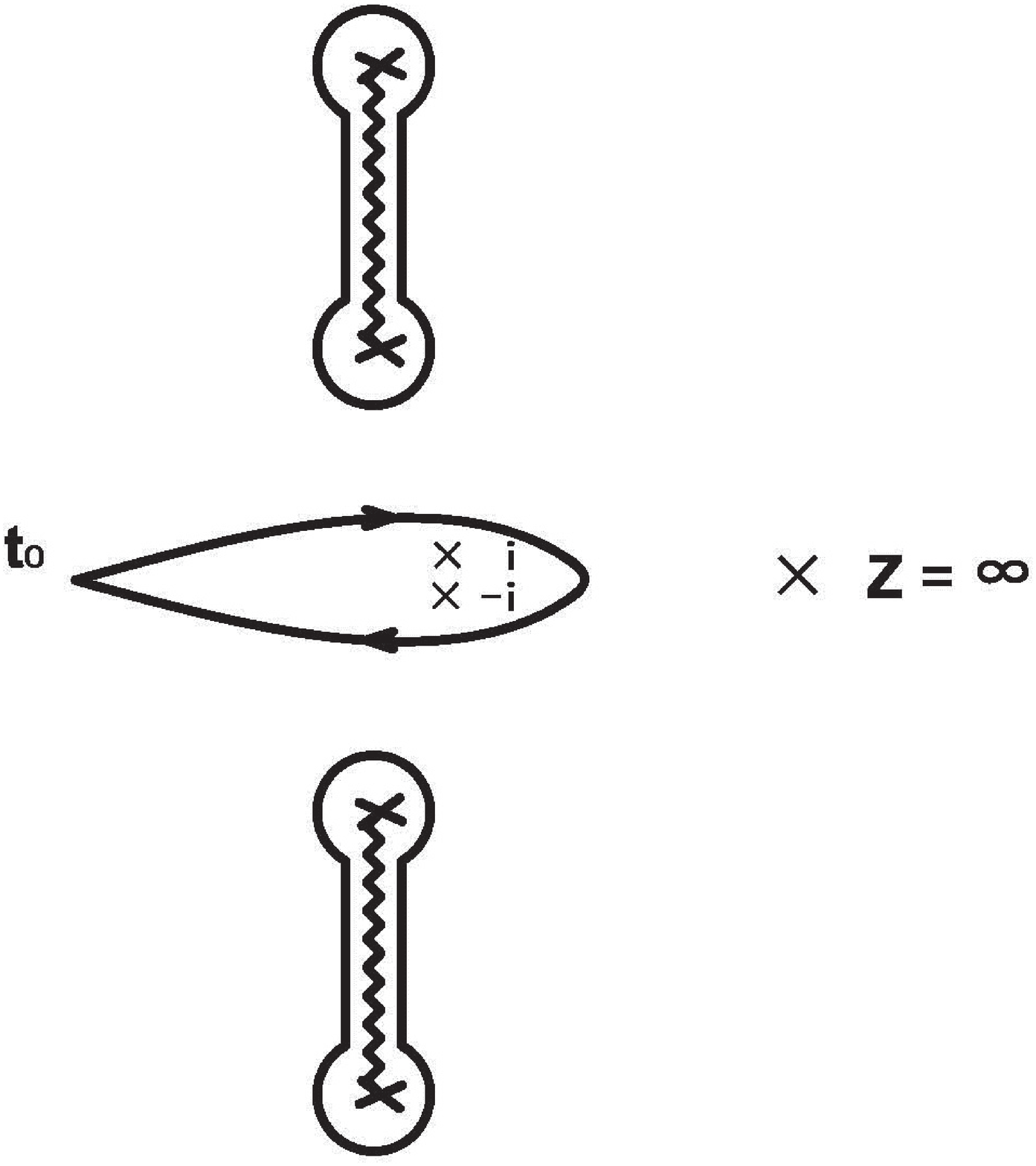}}
\caption{The second class consists of closed paths $C_{II}^{(2)}$ and the third class of closed paths $C_{III}^{(1)}$ that start from $t_0$ and include only one of finite simple poles $z_i = i$ and $z_i = -1$, and they also receive the geometric factor from $z = \infty$ (left panel). The fourth class of closed paths $C_{IV}^{(1)}$ starts from the initial time $t_0$ and includes both simple poles $z_i = i$ and $z_i = -1$, and it also receives the geometric contribution from $z = \infty$ (right pane).} \label{contour-2}
\end{figure}

\section{Conclusion} \label{sec IV}

We have shown that the Stokes phenomenon for dS radiation, constructive interference in even dimensions and destructive interference in odd dimensions, has a geometric interpretation in the complex-time plane. In contrast to the trivial real-time dynamics in the in-in formalism, the transported in-vacuum of a quantum field along a closed path in the complex-time plane may gain geometric contributions from possible simple poles and the magnitude of the scattering amplitude square between the in-vacuum and the transported one gives the particle-production rate for that path. The global coordinates of a dS spacetime have two finite simple poles corresponding to the north and south poles of the Euclidean geometry as well as the simple pole at infinity, corresponding to the cosmological horizon. It is shown that the four classes of paths, either including or not including the finite simple poles, provide each channel for dS radiation, which explains not only the leading Boltzmann factor from the simple pole at infinity but also the sinusoidal behavior from finite simple poles. Thus the Stokes phenomenon for dS radiation has a geometric interpretation in the complex-time plane.

We now compare the geometric interpretation of this paper with other tunneling approaches to dS radiation.
In the tunneling interpretation of dS radiation the cosmological horizon in the static coordinates plays an essential role in emitting particles from vacuum fluctuations near the horizon \cite{Padmanabhan02,Parikh02,Medved02,Zhang05,Volovik,Srinivasan,Kim07,Kim08}.
On the other hand, the geometric interpretation relies on the nonstationary nature of the time-dependent Hamiltonian of a quantum field in dS spacetime.
The quantum evolution in the complex-time plane provides the geometric factor when the in-vacuum is transported along a closed path and returns to the initial time. In fact, the magnitude of the scattering amplitude square between the in-vacuum and the transported one along an independent path gives a channel for particle production. Further, the two simple poles from the north and south poles of the Euclidean geometry result in the interference among independent paths, constructive in even dimensions and destructive in odd dimensions. It would be interesting to investigate physics behind two methods by comparing different coordinates for the embedding spacetime.

\acknowledgments
The author would like to thank Misao Sasaki for useful discussions on de Sitter spacetimes.
He also thanks Eunju Kang for drawing figures. This paper was initiated and completed at Yukawa Institute for Theoretical Physics, Kyoto University.
This work was supported by Basic Science Research Program through
the National Research Foundation of Korea funded by the Ministry of Education (NRF-2012R1A1B3002852).

\end{document}